\documentclass[pra,showpacs,twocolumn]{revtex4}
\usepackage{amssymb}
\usepackage{amsfonts}
\usepackage{amsmath}
\usepackage{graphicx}

\begin{document}

\title{Smoothed quantum-classical states in time-irreversible hybrid dynamics%
}
\author{Adri\'{a}n A. Budini}
\affiliation{Consejo Nacional de Investigaciones Cient\'{\i}ficas y T\'{e}cnicas
(CONICET), Centro At\'{o}mico Bariloche, Avenida E. Bustillo Km 9.5, (8400)
Bariloche, Argentina, and Universidad Tecnol\'{o}gica Nacional (UTN-FRBA),
Fanny Newbery 111, (8400) Bariloche, Argentina}
\date{\today }

\begin{abstract}
We consider a quantum system continuously monitored in time which in turn is
coupled to an arbitrary dissipative classical system (diagonal reduced
density matrix). The quantum and classical dynamics can modify each other,
being described by an arbitrary time-irreversible hybrid Lindblad equation.
Given a measurement trajectory, a conditional bipartite stochastic state can
be inferred by taking into account all previous recording information
(filtering). Here, we demonstrate that the joint quantum-classical state can
also be inferred by taking into account both past and future measurement
results (smoothing). The smoothed hybrid state is estimated without
involving information from unobserved measurement channels. Its average over
recording realizations recovers the joint time-irreversible behavior. As an
application we consider a fluorescent system monitored by an inefficient
photon detector. This feature is taken into account through a fictitious
classical two-level system. The average purity of the smoothed quantum state
increases over that of the (mixed) state obtained from the standard quantum
jump approach.
\end{abstract}

\maketitle

%03.65.Ta Foundations of quantum mechanics, measurement theory
%42.50.Lc Quantum fluctuations, quantum noise, and quantum jumps
%42.50.Dv NonClassical states; quantum state engineering and measurements
%02.50.Tt Inference methods (in probability theory and stochastic processes)
%03.65.Yz Decoherence, open systems; quantum statistical methods
%03.65.Wj State reconstruction, quantum tomography
%03.65.Ca Formalism (in quantum mechanics)

\section{Introduction}

In quantum mechanics the state of a system is described by a state vector,
or more generally, by a density matrix operator in the case of open systems 
\cite{breuerbook}. The environmental influence renders the system dynamics
irreversible in time. In addition, the environment may be continuously
monitored in time by some measuring device. A fundamental problem solved by
the quantum-jump approach \cite{breuerbook,carmichaelbook,plenio,milburn} is
the estimation of the system state conditioned on a given (single)
measurement signal (trajectory). Given the stochastic nature of the
measurement process, the system state inherits this property, while its
average over measurement trajectories recovers the irreversible system
dynamics.

The quantum jump approach delivers a stochastic system state that depends on
all previous measurement results. The estimation is possible after knowing
the system initial condition and its dynamics. Different refinements of this
technique were known in a classical context (classical estimation theory) 
\cite{recipes,jaz}. \textit{Filtering} is a Bayesian estimation technique
where the system state is conditioned on earlier measurements while \textit{%
smoothing} means that both earlier and later measurements are considered.
Hence, the standard quantum jump approach can be considered as a quantum
extension of classical filtering. Different formulations of a
\textquotedblleft quantum version\textquotedblright\ of smoothing have been
achieved recently \cite%
{tsang,tsanPRA,molmer,meschede,murch,haroche,xu,tan,naghi,huard,wiseman}.

The estimation of a \textit{classical parameter} that affect the evolution
of a quantum system using the results of (both earlier and later)
measurements on that system were performed by Tsang in Ref. \cite{tsang}.
Specific physical applications of this approach have been analyzed \cite%
{tsanPRA}. Estimation of the result of a quantum measurement using past and
future information was characterized by Gammelmark, Julsgaard and M\o lmer
in Ref. \cite{molmer}. A \textit{past quantum state}, consisting in a pair
of matrices, density matrix and an \textquotedblleft effect
operator,\textquotedblright\ allows to achieving a better estimation of a
non-selective measurement performed over the system in the past. Extra
analysis and specific implementations were posteriorly characterized \cite%
{meschede,murch,haroche,xu,tan,naghi,huard}. In contrast with the previous
results, a \textit{smoothed quantum state} (positive density operator)
consistent with past and future measurement information was introduced by
Guevara and Wiseman in Ref. \cite{wiseman}. They considered a partially
monitored optical quantum system. The smoothed quantum state can be
estimated after knowing a pure state conditioned on both the observed
(homodyne photocurrent) and unobserved (photon count) records. A significant
recovering of the purity lost due to the unobserved signal is achieved.

Following the previous research lines, in this paper we demonstrate that a
smoothed quantum-classical state can be consistently defined. It describes
the estimated joint state of a dissipative (time-irreversible) hybrid
quantum-classical arrange when past and future measurement signals performed
on the quantum subsystem are taken into account.

In contrast with previous analysis \cite{tsang,molmer}, a joint smoothed
state is explicitly defined. In addition, here the quantum and classical
dynamics are intrinsically entangled (correlated). Each one may modify the
other. Their evolution is described by an arbitrary time-irreversible
(hybrid) Lindblad rate equation \cite{rate,dual}, which correspond to the
more general bipartite evolution restricted by the requirements of
Markovianicity and classicality (one of the reduced density matrixes is
diagonal in a fixed basis at all times). The smoothed joint state can be
estimated after knowing a measurement trajectory, the initials conditions
and the characteristic parameters of the dynamics. By partial trace the
bipartite smoothed state provides the partial (smoothed) states of the
quantum system and the classical counterpart. As in the standard quantum
jump approach, by averaging over realizations (past and futures ones) the
joint irreversible dynamics is recovered.

As an application we consider a single fluorescent system monitored by an
inefficient photon detector. By introducing a fictitious classical degree of
freedom associated to the detector, the formalism applies to this situation.
The purity of the smoothed state increases with respect to that obtained
from the standard quantum jump approach. In contrast with previous
approaches \cite{wiseman}, for estimating the smoothed state it is not
necessary to determine the (past) unobserved signal trajectory. This is a
general feature of the present formalism.

The paper is outlined as follows. In Sec. II we introduce the underlying
formalism that describes the hybrid quantum-classical evolution \cite{rate}.
The corresponding filtered state, equivalent to the quantum jump approach 
\cite{JPB}, is also reviewed. Over this basis, the smoothed
quantum-classical state is developed in Sec. III. As an example, in Sec. IV
the formalism is applied to a fluorescent system monitored by an inefficient
detector. The Conclusions are provided in Sec. IV. Calculus details that
support the main results are provided in the Appendixes.

\section{Quantum-classical dynamics}

We consider a quantum system $S$ interacting with classical
(time-irreversible) degrees of freedom, denoted by $C.$ The bipartite
arrange can be described by an hybrid quantum-classical density operator $%
\left\vert \rho _{t}\right) .$ It is written as%
\begin{equation}
\left\vert \rho _{t}\right) \equiv \sum_{R}(R\left\vert \rho _{t}\right) |R).
\end{equation}%
Here, the index $R$ labels each state of $C,$ which in turn has assigned a
(column) vector $|R)=(0,\cdots 1,\cdots 0)^{\mathrm{T}}.$ The (real)
vectorial base $\{|R)\}$ satisfy $(R|R^{\prime })=\delta _{RR^{\prime }}.$
Each (conditional) density operator $(R\left\vert \rho _{t}\right) $ is
defined in the Hilbert space of $S.$ Introducing the vector $(1|\equiv
\sum_{R}(R|=(1,\cdots ,1),$ the unconditional density operator $\rho _{t}$\
of $S$ follows from the hybrid (vectorial) operator as \cite{rate}%
\begin{equation}
\rho _{t}=(1|\rho _{t})=\sum_{R}(R|\rho _{t}).  \label{RhoS}
\end{equation}%
The vector $\left\vert P_{t}\right) $ of classical probabilities $%
\{P_{t}[R]\}$ for the set of states $\{|R)\}$ can be written as%
\begin{equation}
\left\vert P_{t}\right) =\sum_{R}P_{t}[R]|R)=\sum_{R}\mathrm{Tr}[(R|\rho
_{t})]|R).  \label{Perre}
\end{equation}%
$\mathrm{Tr}[\bullet ]$ denotes trace operation in the Hilbert space of $S.$
With the previous definitions, the vectorial hybrid operator can be
rewritten as%
\begin{equation}
\left\vert \rho _{t}\right) =\sum_{R}P_{t}[R]\frac{(R|\rho _{t})}{\mathrm{Tr}%
[(R|\rho _{t})]}|R).  \label{Vectorial}
\end{equation}%
Hence, $(R|\rho _{t})/\mathrm{Tr}[(R|\rho _{t})]$ is the quantum state of $S$
\textit{given that} $C$ is in the state $|R).$

The more general time-irreversible (Markovian) evolution equation describing
the interaction between $S$ and $C$ is given by a (hybrid) Lindblad rate
equation \cite{rate},%
\begin{equation}
\frac{d\left\vert \rho _{t}\right) }{dt}=\mathcal{\hat{L}}\left\vert \rho
_{t}\right) ,  \label{VectorialLindbladRate}
\end{equation}%
where (separable) initial condition $\left\vert \rho _{0}\right) =\rho
_{0}\left\vert P_{0}\right) $ are assumed.

The (matrix) superoperator $\mathcal{\hat{L}}$ may adopts very different
structures \cite{rate,dual}, which may be used, for example, for describing
radiation patterns in fluorescent systems coupled to classically fluctuating
reservoirs \cite{JPB,OpenSMS} (see also Eq. (\ref{detector}) below). The
following analysis and results are valid for arbitrary $\mathcal{\hat{L}}$\
structures.

\subsection*{Filtered state}

The quantum system $S$ (or equivalently its environment) is continuously
monitored in time. We assume that, up to time $t,$ each recorded measurement
realization consists in a set of random times $\overleftarrow{t}\equiv
\{t_{1},t_{2},\cdots t_{n}\},$ with $0\leq t_{i}\leq t.$ Each time $t_{i}$\
can be associated to the time at which $S$ suffers a given transition. A
filtered state is an estimation $\left\vert \rho _{t}^{\mathrm{st}}\right) $
of the bipartite state conditioned on a given (past) measurement trajectory.
In addition to the times $\overleftarrow{t},$ the initial bipartite state $%
\left\vert \rho _{0}\right) $ and its dynamics [Eq. (\ref%
{VectorialLindbladRate})] are known. The quantum jump approach allows to
define a filtered state $\left\vert \rho _{t}^{\mathrm{st}}\right) ,$ which
relies on the closure condition%
\begin{equation}
\overleftarrow{|\rho _{t}^{\mathrm{st}})}=\left\vert \rho _{t}\right) ,
\label{AveragingPast}
\end{equation}%
that is, the average of $\left\vert \rho _{t}^{\mathrm{st}}\right) $\ over
measurement trajectories (denoted with the over arrow $\leftarrow $)
recovers the dynamics dictated by Eq. (\ref{VectorialLindbladRate}).

The filtered state can be written as (see Appendix A)%
\begin{equation}
\left\vert \rho _{t}^{\mathrm{st}}\right) =\frac{\mathcal{\hat{U}}[t,0,%
\overleftarrow{t}]\left\vert \rho _{0}\right) }{\mathrm{Tr}[(1|\mathcal{\hat{%
U}}[t,0,\overleftarrow{t}]\left\vert \rho _{0}\right) ]}.  \label{Filtered}
\end{equation}%
Here, the (unconditional) vectorial propagator $\mathcal{\hat{U}}$ is
defined as%
\begin{equation}
\mathcal{\hat{U}}[t,t^{\prime },\{\tau _{i}\}_{1}^{n}]\equiv e^{\mathcal{%
\hat{D}}(t-\tau _{n})}\Big{\{}\prod_{i=2}^{n}\mathcal{\hat{J}}e^{\mathcal{%
\hat{D}}(\tau _{i}-\tau _{i-1})}\Big{\}}\mathcal{\hat{J}}e^{\mathcal{\hat{D}}%
(\tau _{1}-t^{\prime })}.  \label{Propa}
\end{equation}%
The vectorial superoperators $\mathcal{\hat{D}}$ and $\mathcal{\hat{J}}$
recover the bipartite dynamics generator%
\begin{equation}
\mathcal{\hat{L}}=\mathcal{\hat{D}}+\mathcal{\hat{J}}.  \label{LDJ}
\end{equation}%
The vectorial superoperator $\mathcal{\hat{J}}$ is chosen such that the
transformation $|\rho )\rightarrow \mathcal{\hat{M}}|\rho ),$\ where%
\begin{equation}
\mathcal{\hat{M}}|\rho )=\frac{\mathcal{\hat{J}}|\rho )}{\mathrm{Tr}[(1|%
\mathcal{\hat{J}}|\rho )]},  \label{EME}
\end{equation}%
corresponds to the measurement transformation of the bipartite state given
that a measurement record occurred. Similarly, a propagator $\mathcal{\hat{T}%
}(t,\tau )$ associated to the superoperator $\mathcal{\hat{D}}=\mathcal{\hat{%
L}}-\mathcal{\hat{J}},$%
\begin{equation}
\mathcal{\hat{T}}(t,\tau )|\rho )\equiv \frac{e^{\mathcal{\hat{D}}(t-\tau
)}|\rho )}{\mathrm{Tr}[(1|e^{\mathcal{\hat{D}}(t-\tau )}\left\vert \rho
\right) ]},  \label{TCond}
\end{equation}%
can be read as the (normalized) transformation of the conditional state
between recording events happening successively at times $\tau $ and $t.$

Given the property defined by Eq. (\ref{AveragingPast}) one can associate a
probability density $P_{t}[\overleftarrow{t}]$ for the occurrence of a given
trajectory (defined by the set of times $\overleftarrow{t}$). It reads%
\begin{equation}
P_{t}[\overleftarrow{t}]=\mathrm{Tr}[(1|\mathcal{\hat{U}}[t,0,\overleftarrow{%
t}]\left\vert \rho _{0}\right) ].  \label{ProbabilityTrajectory}
\end{equation}%
This weight, jointly with the definition (\ref{Filtered}) allow to
demonstrate that the requirement (\ref{AveragingPast}) is in fact fulfilled
(Appendix A).

The joint filtered state $\left\vert \rho _{t}^{\mathrm{st}}\right) ,$ Eq. (%
\ref{Filtered}), through the relations [see Eqs. (\ref{RhoS}) and (\ref%
{Perre})]%
\begin{equation}
\rho _{t}^{\mathrm{st}}=(1|\rho _{t}^{\mathrm{st}}),\ \ \ \ |P_{t}^{\mathrm{%
st}})=\sum_{R}\mathrm{Tr}[(R|\rho _{t}^{\mathrm{st}})]|R),  \label{FilterSC}
\end{equation}%
also allows to estimate the partial states of $S$ and $C.$

\section{Quantum-classical smoothed state}

The state estimation (\ref{Filtered}) is conditioned on measurement results
previous to the time $t,$ that is, the set $\overleftarrow{t}.$ It is also
possible to take into account posterior (future of $t$) recording events up
to a given time $T>t.$ The events between $t$ and $T$ are denoted by $%
\overrightarrow{t}\equiv \{t_{n+1},t_{n+2},\cdots t_{N}\},$ which satisfy $%
t\leq t_{i}\leq T.$ The task now is to find the new estimation for the joint
state (smoothed state) taking into account this extra information.

The joint probability density $P_{T}[\overleftrightarrow{t}]$ for a
trajectory in $(0,T)$\ with detection times $\overleftrightarrow{t}\equiv 
\overleftarrow{t}\cup \overrightarrow{t}=\{t_{i}\}_{1}^{N},$\ from Eq. (\ref%
{ProbabilityTrajectory}), can be written as%
\begin{equation}
P_{T}[\overleftrightarrow{t}]=\mathrm{Tr}[(1|\mathcal{\hat{U}}[T,t,%
\overrightarrow{t}]\mathcal{\hat{U}}[t,0,\overleftarrow{t}]\left\vert \rho
_{0}\right) ].  \label{PT}
\end{equation}%
This object can also be expressed as%
\begin{equation}
P_{T}[\overleftrightarrow{t}]=\sum_{R_{t}}P_{T}[\overleftrightarrow{t}%
,R_{t}].  \label{norma}
\end{equation}%
Here, $P_{T}[\overleftrightarrow{t},R_{t}]$ is the joint probability density
of the random variables $\overleftrightarrow{t}$ and $R_{t}.$ The last one
labels the state of the classical degrees of freedom at time $t.$ From Eq. (%
\ref{PT}) we write%
\begin{equation}
P_{T}[\overleftrightarrow{t},R_{t}]=\mathrm{Tr}[(1|\mathcal{\hat{U}}[T,t,%
\overrightarrow{t}]|R_{t})(R_{t}|\mathcal{\hat{U}}[t,0,\overleftarrow{t}%
]\left\vert \rho _{0}\right) ].  \label{PtR}
\end{equation}%
This expression can be interpreted in terms of a (classical) measurement
performed over $C$ at time $t.$ Using that $\sum_{R_{t}}|R_{t})(R_{t}|$ is
the identity matrix in the vectorial space of $C,$ it follows that the
normalization (\ref{norma}) is satisfied trivially.

By introducing the conditional probability $P_{T}[R_{t}|\overleftrightarrow{t%
}]$ of $R_{t}$ \textit{given} the set $\overleftrightarrow{t},$ Bayes rule
gives the relation%
\begin{equation}
P_{T}[\overleftrightarrow{t},R_{t}]=P_{T}[R_{t}|\overleftrightarrow{t}]P_{T}[%
\overleftrightarrow{t}].
\end{equation}%
Hence, from Eqs. (\ref{norma}) and (\ref{PtR}) we get%
\begin{equation}
P_{T}[R_{t}|\overleftrightarrow{t}]=\frac{\mathrm{Tr}[(1|\mathcal{\hat{U}}%
[T,t,\overrightarrow{t}]|R_{t})(R_{t}|\mathcal{\hat{U}}[t,0,\overleftarrow{t}%
]\left\vert \rho _{0}\right) ]}{\sum_{R}\mathrm{Tr}[(1|\mathcal{\hat{U}}[T,t,%
\overrightarrow{t}]|R)(R|\mathcal{\hat{U}}[t,0,\overleftarrow{t}]\left\vert
\rho _{0}\right) ]}.  \label{PSmoothErre}
\end{equation}%
This expression allow us to estimate the state of the classical degrees of
freedom $C$ at time $t$ given that we know both past and future measurement
results $(\overleftrightarrow{t}=\overleftarrow{t}\cup \overrightarrow{t})$
performed on the quantum system $S$ in the time interval $(0,T).$

Now, we introduce the unnormalized joint filtered state%
\begin{equation}
|\tilde{\rho}_{t}^{\mathrm{st}})\equiv \mathcal{\hat{U}}[t,0,\overleftarrow{t%
}]\left\vert \rho _{0}\right) ,
\end{equation}%
and the \textquotedblleft effect vectorial-operator\textquotedblright 
\begin{equation}
|E_{t}^{\mathrm{st}})\equiv \mathcal{\hat{U}}^{\#}[T,t,\overrightarrow{t}]|%
\mathrm{I}).
\end{equation}%
Here $|\mathrm{I})\equiv \mathrm{I}|1),$ where $\mathrm{I}$\ is the identity
matrix in the Hilbert space of $S.$ Furthermore, $\mathcal{\hat{U}}^{\#}$ is
the dual propagator of $\mathcal{\hat{U}}.$ It is defined by the relation 
\cite{dual}%
\begin{equation}
\mathrm{Tr}[(A|\mathcal{\hat{U}}|\rho )]=\mathrm{Tr}[(\rho |\mathcal{\hat{U}}%
^{\#}|A)],
\end{equation}%
where $|\rho )$ and $|A)$ are arbitrary vectorial states and operators
respectively. Using that $\mathrm{Tr}[(A|\mathcal{\hat{U}\hat{V}}|\rho )]=%
\mathrm{Tr}[(\rho |\mathcal{\hat{V}}^{\#}\mathcal{\hat{U}}^{\#}|A)],$ from
Eq. (\ref{Propa}) it follows%
\begin{eqnarray}
\mathcal{\hat{U}}^{\#}[t,t^{\prime },\{\tau _{i}\}_{1}^{n}] &=&e^{\mathcal{%
\hat{D}}^{\#}(\tau _{1}-t^{\prime })}\mathcal{\hat{J}}^{\#}\Big{\{}%
\prod_{i=2}^{n}e^{\mathcal{\hat{D}}^{\#}(\tau _{i}-\tau _{i-1})}\mathcal{%
\hat{J}}^{\#}\Big{\}}  \notag \\
&&\times e^{\mathcal{\hat{D}}^{\#}(t-\tau _{n})},  \label{Dual}
\end{eqnarray}%
where $\mathcal{\hat{D}}^{\#}$ and $\mathcal{\hat{J}}^{\#}$\ are the dual
operators to $\mathcal{\hat{D}}$ and $\mathcal{\hat{J}}$ respectively (see
Ref. \cite{dual}).

Given that%
\begin{equation}
\mathrm{Tr}[(1|\mathcal{\hat{U}}[T,t,\overrightarrow{t}]|\rho )]=\mathrm{Tr}%
[(\rho |\mathcal{\hat{U}}^{\#}[T,t,\overrightarrow{t}]|\mathrm{I})],
\end{equation}%
the probability (\ref{PSmoothErre}) can be written in terms of $|\tilde{\rho}%
_{t}^{\mathrm{st}})$\ and $|E_{t}^{\mathrm{st}})$ as%
\begin{equation}
P_{T}[R_{t}|\overleftrightarrow{t}]=\frac{\mathrm{Tr}[(\tilde{\rho}_{t}^{%
\mathrm{st}}|R_{t})(R_{t}\left\vert E_{t}^{\mathrm{st}}\right) ]}{\sum_{R}%
\mathrm{Tr}[(\tilde{\rho}_{t}^{\mathrm{st}}|R)(R\left\vert E_{t}^{\mathrm{st}%
}\right) ]}.  \label{PastState}
\end{equation}%
The structure of this equation is similar to that obtained in Refs. \cite%
{tsang} and \cite{molmer}, where the pair $\{|\tilde{\rho}_{t}^{\mathrm{st}%
}),\left\vert E_{t}^{\mathrm{st}}\right) \}$ can be named as a
\textquotedblleft vectorial past quantum state.\textquotedblright\
Trivially, under the replacement $|\tilde{\rho}_{t}^{\mathrm{st}%
})\rightarrow |\rho _{t}^{\mathrm{st}}),$ the smoothed probability $%
P_{T}[R_{t}|\overleftrightarrow{t}]$ can also be written in terms of the
normalized filtered state (\ref{Filtered}). Furthermore, for $T\rightarrow t$
(filtering), Eq. (\ref{FilterSC}) is recovered, $\lim_{T\rightarrow
t}P_{T}[R_{t}|\overleftrightarrow{t}]=P_{t}[R_{t}|\overleftarrow{t}]=\mathrm{%
Tr}[(\rho _{t}^{\mathrm{st}}|R_{t})].$

From $P_{T}[R_{t}|\overleftrightarrow{t}]$ it is possible to define a
smoothed quantum-classical state $|\rho _{t,T}^{\mathrm{st}}),$ that is, an
estimation of the quantum-classical joint state taking into account both
past and future measurement results. From Eq.~(\ref{Vectorial}), we write%
\begin{equation}
|\rho _{t,T}^{\mathrm{st}})=\sum_{R}P_{T}[R|\overleftrightarrow{t}]\frac{%
(R|\rho _{t}^{\mathrm{st}})}{\mathrm{Tr}[(R|\rho _{t}^{\mathrm{st}})]}|R),
\label{Main}
\end{equation}%
where $|\rho _{t}^{\mathrm{st}})$ is the filtered state defined in Eq. (\ref%
{Filtered}) while $P_{T}[R|\overleftrightarrow{t}]$ follows from Eq. (\ref%
{PastState}). This is the\ main result of this section. Notice that $|\rho
_{t,T}^{\mathrm{st}})$ can be obtained after knowing the measurement
results, the joint initial state and the quantum-classical dynamics [see
Eqs. (\ref{Propa}) and (\ref{Dual})].

The previous result relies on the fact that the state of $S$ \textit{given}
that $C$ is in the state $|R)$ at time $t$ is given by $(R|\rho _{t}^{%
\mathrm{st}})/\mathrm{Tr}[(R|\rho _{t}^{\mathrm{st}})].$ Hence, the smoothed
probability $P_{T}[R|\overleftrightarrow{t}]$ is the correct weight of each
contribution given that we know both past and future measurement results.

Similarly to the filtering case [Eq. (\ref{FilterSC})], the relations%
\begin{equation}
\rho _{t,T}^{\mathrm{st}}=(1|\rho _{t,T}^{\mathrm{st}}),\ \ \ \ \ \ \
|P_{t,T}^{\mathrm{st}})=\sum_{R}\mathrm{Tr}[(R|\rho _{t,T}^{\mathrm{st}})]\
|R),  \label{PartialSmooth}
\end{equation}%
correspond to the smoothed estimations of the partial states of $S$ and $C$
respectively.

In Appendix B we demonstrate that by averaging the smoothed joint state $%
|\rho _{t,T}^{\mathrm{st}})$ over future measurement results, $|\rho _{t,T}^{%
\mathrm{st}})\rightarrow \overrightarrow{|\rho _{t,T}^{\mathrm{st}})},$ the
filtered state $|\rho _{t}^{\mathrm{st}})$ is recovered,%
\begin{equation}
\overrightarrow{|\rho _{t,T}^{\mathrm{st}})}=|\rho _{t}^{\mathrm{st}}),\ \ \
\ \ \ \ \Rightarrow \ \ \ \ \ \ \ \ \ \overleftrightarrow{|\rho _{t,T}^{%
\mathrm{st}})}=\left\vert \rho _{t}\right) .  \label{Promedios}
\end{equation}%
In addition, the second equality follows straightforwardly from the former
one and Eq. (\ref{AveragingPast}). The over arrow $\leftrightarrow $ means
average over both past and future measurement results. Thus, the average of
both the joint smoothed and filtered states recover the irreversible
dynamics of the quantum-classical arrange, Eq. (\ref{VectorialLindbladRate}).

\section{Inefficient photon-detection}

The formalism developed in the previous section may have applications in
different contexts. For example, the dynamics of fluorescent systems coupled
to classically self-fluctuating reservoirs \cite{JPB,OpenSMS} can be
described through different Lindblad rate equations. The classical degrees
of freedom correspond to different configurational states of the
environment. In addition, as the formalism can be applied independently of
the physical origin of the classical counterpart, here we apply the previous
results to a different physical situation.

We consider a single fluorescent two-level system (with states $|\pm \rangle 
$) coupled to a resonant laser field. The system-laser (time-reversible)
coupling is proportional to Rabi frequency $\Omega ,$ while its natural
(time-irreversible) decay rate is $\gamma .$ The evolution of its density
matrix $\rho _{t}$ is \cite{breuerbook,milburn}%
\begin{equation}
\dfrac{d\rho _{t}}{dt}=-\dfrac{i\Omega }{2}[\sigma _{x},\rho
_{t}]_{-}+\gamma (\sigma \rho _{t}\sigma ^{\dagger }-\{\sigma ^{\dagger
}\sigma ,\rho _{t}\}_{+}),  \label{fluor}
\end{equation}%
where $\sigma _{x}$ is the $x$-Pauli matrix, while $\sigma =|-\rangle
\langle +|,$ and $\sigma ^{\dagger }=|-\rangle \langle +|.$ Furthermore, $%
[p,q]_{-}\equiv pq-qp,$ while $\{p,q\}_{+}\equiv (pq+qp)/2$ denotes an
anticommutator.

The scattered radiation field is measured by an inefficient photon detector
whose efficiency is $\eta .$ The standard quantum jump approach covers this
situation \cite{milburn}. Its description can be recovered from Sec. II in
the limit in which the classical system $C$ has only one state. The
splitting defined by Eq. (\ref{LDJ}) is performed by introducing the
(unidimensional) superoperators \cite{milburn}%
\begin{equation}
\mathcal{J}[\rho ]=\gamma \eta \sigma \rho \sigma ^{\dagger },\ \ \ \ \ \ \
\ \ \mathcal{D}=\mathcal{L}-\mathcal{J},  \label{JDFluor}
\end{equation}%
where $\mathcal{L}$ follows from Eq. (\ref{fluor}), $d\rho _{t}/dt=\mathcal{L%
}[\rho ].$ Given that%
\begin{equation}
\mathcal{M}[\rho ]=\frac{\mathcal{J}[\rho ]}{\mathrm{Tr}[\mathcal{J}\rho ]}%
=|-\rangle \langle -|,  \label{MFluor}
\end{equation}%
the system resets to its ground state in each detection event. In
consequence the emission process is a renewal one. A waiting time
distribution \cite{carmichaelbook} gives the probability density of the time
interval between consecutive events (see Appendix C). The filtered state $%
\rho _{t}^{\mathrm{st}}$ [Eq. (\ref{Filtered})] is not pure, $1/2\leq 
\mathrm{Tr}[(\rho _{t}^{\mathrm{st}})^{2}]\leq 1.$ Nevertheless, for $\eta
=1 $ (perfect detector), a pure state is obtained, $\mathrm{Tr}[(\rho _{t}^{%
\mathrm{st}})^{2}]=1$ (strictly, this equality is valid in general after the
first detection event). Our goal here is to get a new estimation of the
system state using the smoothing technique described previously.

\subsection{Quantum-classical representation}

The measurement trajectory is given by the detection times obtained from the
inefficient detector. Clearly, the system [Eq. (\ref{fluor})] does not
include any classical degree of freedom. Nevertheless, one can introduce a
fictitious classical system that takes into account the imperfection of the
detector while the (quantum) system dynamics remains the same. It is
described by two (classical) states denoted by $|a),$ with $a=d$ (detected)
and $a=u$ (undetected) (Fig. 1). The joint vectorial state $|\rho _{t})$ is
defined by the matrixes $(a|\rho _{t})=\rho _{t}^{a}.$ Their evolution is
given by the Lindblad rate equation%
\begin{eqnarray}
\dfrac{d\rho _{t}^{a}}{dt} &=&-\dfrac{i\Omega }{2}[\sigma _{x},\rho
_{t}^{a}]_{-}+\gamma _{a}(\sigma \rho _{t}^{a}\sigma ^{\dagger }-\{\sigma
^{\dagger }\sigma ,\rho _{t}^{a}\}_{+})  \notag \\
&&-\gamma _{b}\{\sigma ^{\dagger }\sigma ,\rho _{t}^{a}\}_{+}+\gamma
_{a}\sigma \rho _{t}^{b}\sigma ^{\dagger },  \label{detector}
\end{eqnarray}%
where the indexes are $a=d,u$ while $b=u,d.$ The decay and coupling rates are%
\begin{equation}
\gamma _{d}\equiv \gamma \eta ,\ \ \ \ \ \ \ \ \gamma _{u}\equiv \gamma
(1-\eta ).
\end{equation}%
The initial conditions are taken as $\rho _{0}^{d}=|-\rangle \langle -|,$
and $\rho _{0}^{u}=~0.$ 
%figura1%figura%figura%figura%figurav%figura%figura%figura%figura%figura%figura%figura%figura%figura%figurav%figura%figura%figura%figura%figura
%figura%figura%figura%figura%figurav%figura%figura%figura%figura%figura%figura%figura%figura%figura%figurav%figura%figura%figura%figura%figura
\begin{figure}[tbp]
\includegraphics[bb=105 70 595
285,angle=0,width=7.cm]{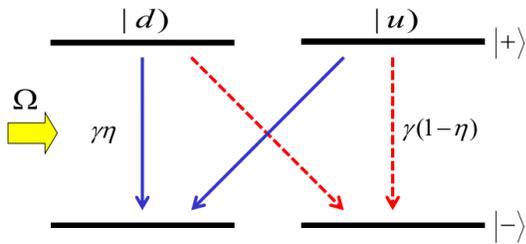}
\caption{Scheme levels corresponding to the evolution (\protect\ref{detector}%
). The quantum system is characterized by the states $|\pm \rangle ,$ while
the classical one by the states $|d)$ and $|u).$ The transition rates are $%
\protect\gamma \protect\eta $ (blue full lines) and $\protect\gamma (1-%
\protect\eta )$ (red dotted lines). The quantum system is coupled to an
external laser field with Rabi frequency $\Omega .$}
\end{figure}
%figura%figura%figura%figura%figurav%figura%figura%figura%figura%figura%figura%figura%figura%figura%figurav%figura%figura%figura%figura%figura
%figura%figura%figura%figura%figurav%figura%figura%figura%figura%figura%figura%figura%figura%figura%figurav%figura%figura%figura%figura%figura

The evolution (\ref{detector}) can be read as follows (see Fig. 1). The
quantum system can suffers the transition $|+\rangle \rightarrow |-\rangle $
with rates $\gamma \eta $ and $\gamma (1-\eta )$ when the classical system
is in the states $|d)$ and $|u)$ respectively. In addition, the transitions $%
|+\rangle |d)\rightarrow |-\rangle |u)$ and $|+\rangle |u)\rightarrow
|-\rangle |d)$ happen with rates $\gamma (1-\eta )$ and $\gamma \eta $
respectively. Therefore, transitions with rate $\gamma \eta $ (detected
events) collapse $C$ to the state $|d),$ while transitions with rate $\gamma
(1-\eta )$ (undetected events) collapse $C$ to the state $|u).$
Independently of the state of the classical system, the fluorescent one is
coupled to the external laser with Rabi frequency $\Omega .$

Given the hybrid evolution (\ref{detector}), the system dynamics follows
from Eq. (\ref{RhoS}), $\rho _{t}=\rho _{t}^{d}+\rho _{t}^{u}.$ It is simple
to check that $(d/dt)\rho _{t}$ obtained in this way obeys the Lindblad
evolution (\ref{fluor}), while the previous initial conditions implies that $%
\rho _{0}=\rho _{0}^{d}+\rho _{0}^{u}=|-\rangle \langle -|.$ Therefore, the
fictitious classical degrees of freedom $C$ associated to the Lindblad rate
equation (\ref{detector}) do not affect the dynamics of the quantum system $S
$ defined by Eq. (\ref{fluor}). We remark that this property is valid for
any value of the characteristic parameters $\Omega ,$ $\gamma ,$ and $\eta .$

The dynamics of $C$ is strongly correlated with the behavior of $S.$ It
starts in the state $|d).$ Transitions between its states $%
|d)\leftrightarrow |u)$ only may happen when $S$ is in the upper state $%
|+\rangle .$ For $\eta =1$ it remains in the initial condition, that is, the
state $|d).$

\subsection{Quantum-classical filtered state}

The filtered joint state $\left\vert \rho _{t}^{\mathrm{st}}\right) $ [Eq. (%
\ref{Filtered})] can be obtained after defining the splitting (\ref{LDJ}).
We choose the vectorial superoperator $\mathcal{\hat{J}}$ such that $%
\mathcal{\hat{M}}$ represents the measurement transformation corresponding
to the \textit{detected photons}, that is, the transitions with rate $\gamma
\eta $ in Fig. 1. Therefore, $[|\rho )=\rho ^{d}|d)+\rho ^{u}|u)]$%
\begin{equation}
\mathcal{\hat{J}}|\rho )=\gamma _{d}[\sigma \rho ^{d}\sigma ^{\dagger
}+\sigma \rho ^{u}\sigma ^{\dagger }]|d),  \label{Jvectorial}
\end{equation}%
which in explicit form reads $\mathcal{\hat{J}}|\rho )=\gamma _{d}(\langle
+|\rho ^{d}|+\rangle +\langle +|\rho ^{u}|+\rangle )|-\rangle \langle -||d).$
The measurement transformation [Eq. (\ref{EME})] becomes%
\begin{equation}
\mathcal{\hat{M}}|\rho )=|-\rangle \langle -||d).  \label{EMEVectorial}
\end{equation}%
Hence, $S$ and $C$ are reseted to the states $|-\rangle \langle -|$ and $|d)$
respectively. On the other hand, the conditional evolution (\ref{TCond}) is
defined with $\mathcal{\hat{D}}=\mathcal{\hat{L}}-\mathcal{\hat{J}},$ where $%
\mathcal{\hat{L}}$ follows from the Lindblad rate equation (\ref{detector}).
With these definitions $(\mathcal{\hat{J}}$ and $\mathcal{\hat{D}}),$ the
joint filtered state $\left\vert \rho _{t}^{\mathrm{st}}\right) $ [Eq. (\ref%
{Filtered})] can be determine after knowing the times of the detection
events.

In a realistic experimental situation, the measurement trajectory is
provided by the (inefficient) photon detector. Here, they are numerically
implemented from a waiting time distribution \cite{carmichaelbook}\ that
gives the probability density for the time intervals between consecutive
events (Appendix C).

In Fig.~2 we show a realization of the $S$ and $C$ filtered states (light
blue lines) [Eq. (\ref{FilterSC})] through the upper system population $%
\langle +|\rho _{t}^{\mathrm{st}}|+\rangle $ and the population $(d|P_{t}^{%
\mathrm{st}}).$ The disruptive changes in the states are associated to the
detection times. Furthermore, the corresponding purities are also shown, $%
\mathrm{Tr}[(\rho _{t}^{\mathrm{st}})^{2}]$ and $purity(d,u)=(d|P_{t}^{%
\mathrm{st}})^{2}+(u|P_{t}^{\mathrm{st}})^{2}.$ For a perfect detector $\eta
=1,$ at any time these last two objects are equal to one.%
%figura1%figura%figura%figura%figurav%figura%figura%figura%figura%figura%figura%figura%figura%figura%figurav%figura%figura%figura%figura%figura
%figura%figura%figura%figura%figurav%figura%figura%figura%figura%figura%figura%figura%figura%figura%figurav%figura%figura%figura%figura%figura
\begin{figure}[tbp]
\includegraphics[bb=35 590 725 1145,angle=0,width=8.5cm]{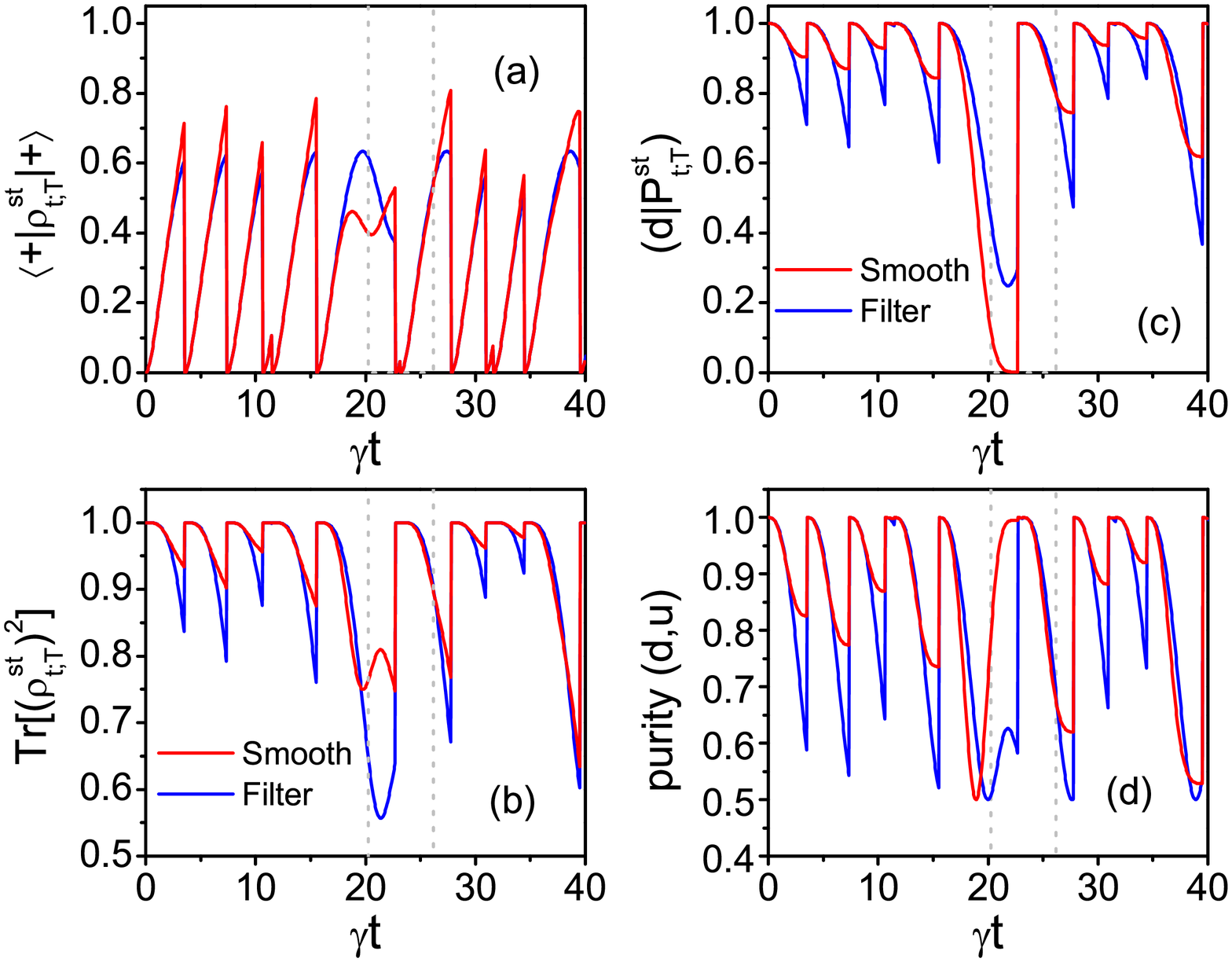}
\caption{Realizations for filtered (light blue lines) and smooth (dark red
lines) states associated to Eq. (\protect\ref{detector}). (a) Upper system
population $\langle +|\protect\rho _{t,T}^{\mathrm{st}}|+\rangle .$ (b)
System purity $\mathrm{Tr}[(\protect\rho _{t,T}^{\mathrm{st}})^{2}].$ (c)
Classical population $(d|P_{t,T}^{\mathrm{st}}).$ (d) $purity(d,u)\equiv
(d|P_{t,T}^{\mathrm{st}})^{2}+(u|P_{t,t}^{\mathrm{st}})^{2}.$ The filtered
states correspond to $T=t.$ In all cases, the parameter are $\Omega /\protect%
\gamma =1$ and $\protect\eta =0.8.$ For the smoothed realizations, $T$ is
chosen such that $\protect\gamma (T-t)=30.$ The vertical dashed lines are
the times of the undetected events.}
\end{figure}
%figura%figura%figura%figura%figurav%figura%figura%figura%figura%figura%figura%figura%figura%figura%figurav%figura%figura%figura%figura%figura
%figura%figura%figura%figura%figurav%figura%figura%figura%figura%figura%figura%figura%figura%figura%figurav%figura%figura%figura%figura%figura

A fundamental property of the quantum system realizations shown in Fig. 2 is
that they are exactly the same than those obtained from the standard quantum
jump approach applied to the Lindblad evolution (\ref{fluor}). In fact, it
is straightforward to demonstrate that%
\begin{equation}
(1|\mathcal{\hat{D}}|\rho )=\mathcal{D}(1|\rho )=\mathcal{D}(\rho ^{d}+\rho
^{u}),
\end{equation}%
where $|\rho )=\rho ^{d}|d)+\rho ^{u}|u)$ and $\mathcal{D}$ is defined by
Eq. (\ref{JDFluor}). Given that $\rho ^{d}+\rho ^{u}$ gives the state of $S,$
and given that $(1|\mathcal{\hat{M}}|\rho )=|-\rangle \langle -|$ [Eq. (\ref%
{EMEVectorial})] it follows that the realizations of $(1|\rho _{t}^{\mathrm{%
st}})=\rho _{t}^{\mathrm{st}}$ coincide with the realizations of the
standard quantum jump approach defined from the measurement superoperator (%
\ref{JDFluor}) and (\ref{MFluor}). The measurement statistics is also the
same (Appendix C). Thus, not only the irreversible evolution of the $S,$ but
also the filtered state obtained from the Lindblad rate equation (\ref%
{detector}) is the same than that obtained from Eq. (\ref{fluor}). This is
the main property that sustains the ansatz given by Eq. (\ref{detector}).

\subsection{Quantum-classical smoothed state}

The representation of the fluorescent system monitored by an inefficient
detector in terms of a Lindblad rate equation allows us to define a joint
smoothed state. In fact, given the detection times, it follows from Eq. (\ref%
{Main}), while the partial states from Eq. (\ref{PartialSmooth}).

In Fig. 2, for the same realization of measurement events, we also plot the
smoothed states through the upper system population $\langle +|\rho _{t,T}^{%
\mathrm{st}}|+\rangle $ and the population $(d|P_{t,T}^{\mathrm{st}}).$ The
plotted purities are $\mathrm{Tr}[(\rho _{t,T}^{\mathrm{st}})^{2}]$ and $%
purity(d,u)=(d|P_{t,T}^{\mathrm{st}})^{2}+(u|P_{t,T}^{\mathrm{st}})^{2}.$

The filtered and smoothed realizations develop disruptive events at the same
(detection) times. Nevertheless, for both $S$ and $C,$ the smoothed purities
are higher than the filtered purities. The increment of the smoothed
purities is a consequence of the general result (\ref{Promedios}), that is,
averaging the smoothed states over future measurements events one recover
the filtered states.

In an experimental situation it is impossible to determine when the detector
fails. Nevertheless, given that here we are determining the measurement
events in a numerical way (see Appendix C), it is possible to know when the
undetected events happen. In Fig. 2 they are indicated by the vertical
dashed lines. These times are not necessary for defining the joint smoothed
state (\ref{Main}). Nevertheless, they allow us to understand some features
of the smoothed states. While $\langle +|\rho _{t,T}^{\mathrm{st}}|+\rangle $
does not develops any special characteristic, around $\gamma t\simeq 20$ the
smoothed realization of $(d|P_{t,T}^{\mathrm{st}})$ anticipates the behavior
of the filtered population $(d|P_{t}^{\mathrm{st}}).$ This signature is also
observed in other quantum optical arranges \cite{wiseman}. In addition, here
the smoothed realization almost vanishes, property consistent with the fact
that undetected measurement events leads to the (unobserved) transition $%
|P)\rightarrow |u)$ (see Fig. 1).

The recovering of the purity lost due to the inefficient detector can be
quantified by averaging over an ensemble of measurement events. In Fig. 3 we
plot the smoothed and filtered averaged purities of $S,$ $%
\overleftrightarrow{\mathrm{Tr}[(\rho _{t,T}^{\mathrm{st}})^{2}]}$ and $%
\overleftarrow{\mathrm{Tr}[(\rho _{t}^{\mathrm{st}})^{2}]}$ respectively.
The plots were obtained by averaging $5\times 10^{3}$ realizations. Under
smoothing, with $\eta =0.8,$ about $10\%$ of the purity lost is recovered
when compared with the filtered purity. For $\eta =0.9$ the recovering is
around $15\%.$ These results are similar to that obtained in Ref. \cite%
{wiseman} with a different measurement arrange.

In Fig. 3 it is also shown the filtered and smoothed purities of $C.$ While
these objects refer to the fictitious classical system associated to the
imperfection of the detector, the graphics consistently show similar
properties to that of the quantum counterpart $S.$ 
%figura1%figura%figura%figura%figurav%figura%figura%figura%figura%figura%figura%figura%figura%figura%figurav%figura%figura%figura%figura%figura
%figura%figura%figura%figura%figurav%figura%figura%figura%figura%figura%figura%figura%figura%figura%figurav%figura%figura%figura%figura%figura
\begin{figure}[tbp]
\includegraphics[bb=35 590 725 1145,angle=0,width=8.5cm]{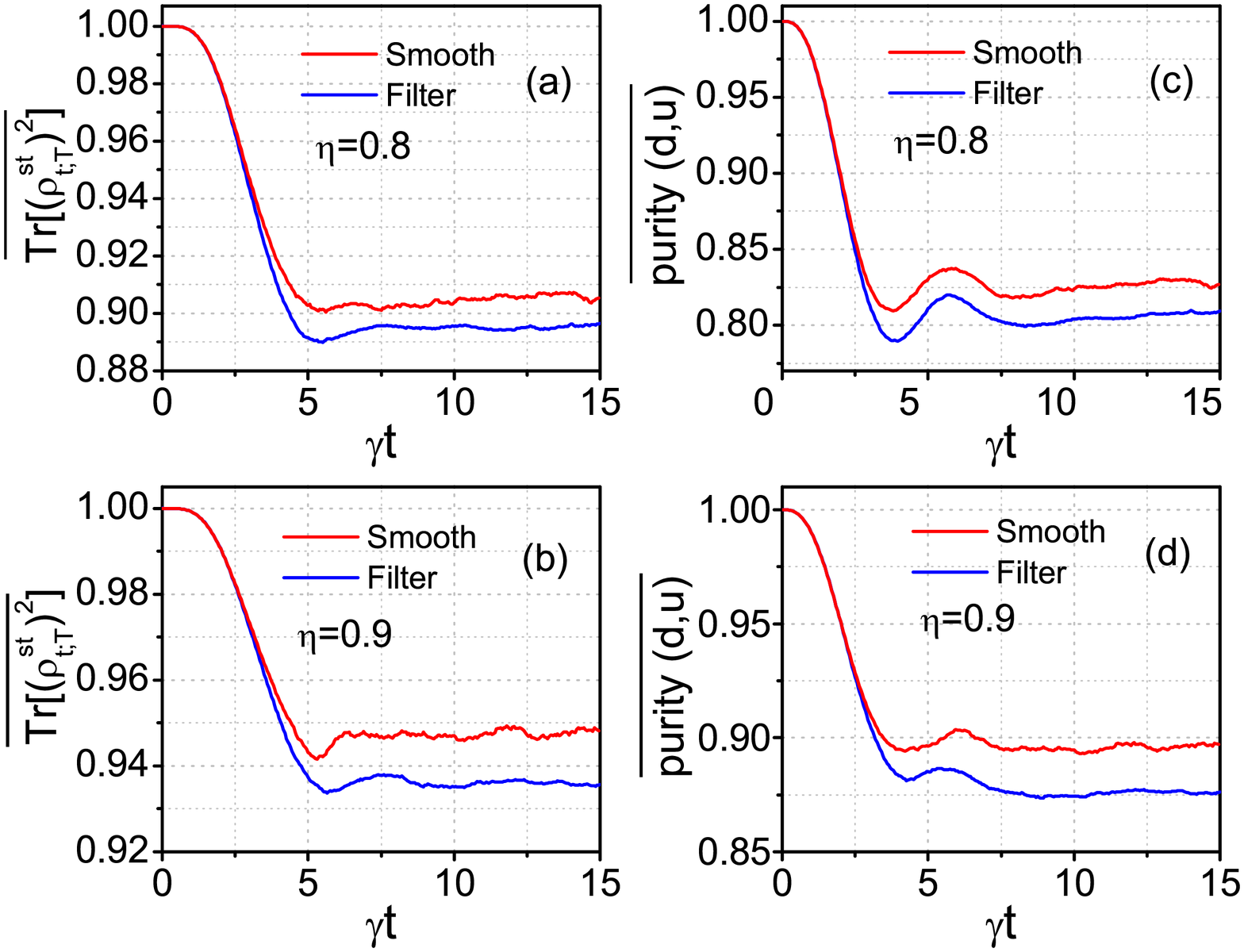}
\caption{Purities of the smoothed and filtered partial states when averaged
over $5\times 10^{3}$ realizations (see Fig. 2). (a) and (b) correspond to $%
\protect\overleftrightarrow{\mathrm{Tr}[(\protect\rho _{t,T}^{\mathrm{st}%
})^{2}]}$ (smoothed) and $\protect\overleftarrow{\mathrm{Tr}[(\protect\rho %
_{t}^{\mathrm{st}})^{2}]}$ (filtered). (c) and (d) to $\protect%
\overleftrightarrow{(d|\protect\rho _{t,T}^{\mathrm{st}})^{2}+(u|\protect%
\rho _{t,T}^{\mathrm{st}})^{2}}$ (smoothed) and $\protect\overleftarrow{(d|%
\protect\rho _{t}^{\mathrm{st}})^{2}+(u|\protect\rho _{t}^{\mathrm{st}})^{2}}
$ (filtered), both objects being denoted as $\overline{purity(d,u)}.$ The
parameters are $\Omega /\protect\gamma =1,$ while $\protect\eta $ is
indicated in each plot.}
\end{figure}
%figura%figura%figura%figura%figurav%figura%figura%figura%figura%figura%figura%figura%figura%figura%figurav%figura%figura%figura%figura%figura
%figura%figura%figura%figura%figurav%figura%figura%figura%figura%figura%figura%figura%figura%figura%figurav%figura%figura%figura%figura%figura

\subsection{Ensemble behavior}

When averaged over an ensemble of realizations [Eq. (\ref{AveragingPast})
and (\ref{Promedios})] both the joint filtered and smoothed states must to
recover the dynamics given by the Lindblad rate equation (\ref{detector}).
For the quantum subsystem, the evolution is exactly the same as that
obtained from the standard Lindblad equation (\ref{fluor}).

In order to check these properties, in Fig. 4 we plot the averaged smoothed$%
\ \overleftrightarrow{\langle +|\rho _{t,T}^{\mathrm{st}}|+\rangle }$ and
filtered $\overleftarrow{\langle +|\rho _{t}^{\mathrm{st}}|+\rangle }$
system populations. Consistently, both averages recover the analytical
solution $\langle +|\rho _{t}|+\rangle $ that follows from Eq. (\ref{fluor})
[idem Eq. (\ref{detector})], which is independent of $\eta .$ The curves are
indistinguishable in the scale of the plots. The same property is valid for
the smoothed $\overleftrightarrow{(d|P_{t,T}^{\mathrm{st}})}$ and filtered $%
\overleftarrow{(d|P_{t}^{\mathrm{st}})}$ averaged classical populations. The
analytical solution of these objects follows from Eq.~(\ref{detector}).

From Eq. (\ref{detector}), it is simple to check that $\lim_{t\rightarrow
\infty }(d|\rho _{t})=\eta \rho _{\infty }$ and $\lim_{t\rightarrow \infty
}(u|\rho _{t})=(1-\eta )\rho _{\infty }$ where $\rho _{\infty
}=\lim_{t\rightarrow \infty }\rho _{t}$ is the stationary solution of Eq. (%
\ref{fluor}), $\rho _{\infty }=\{\{\Omega ^{2},-i\gamma \Omega \},\{i\gamma
\Omega ,\gamma ^{2}+\Omega ^{2}\}\}/(\gamma ^{2}+2\Omega ^{2}).$ In Fig. 4,
the (analytical) stationary values $\lim_{t\rightarrow \infty }\langle
+|\rho _{t}|+\rangle =\Omega ^{2}/(\gamma ^{2}+2\Omega ^{2})=1/3,$ and $%
\lim_{t\rightarrow \infty }(d|P_{t})=\eta =0.8$ are also correctly achieved.%
%
%
%
%figura1%figura%figura%figura%figurav%figura%figura%figura%figura%figura%figura%figura%figura%figura%figurav%figura%figura%figura%figura%figura
%figura%figura%figura%figura%figurav%figura%figura%figura%figura%figura%figura%figura%figura%figura%figurav%figura%figura%figura%figura%figura
\begin{figure}[tbp]
\includegraphics[bb=37 864 722 1133,angle=0,width=8.5cm]{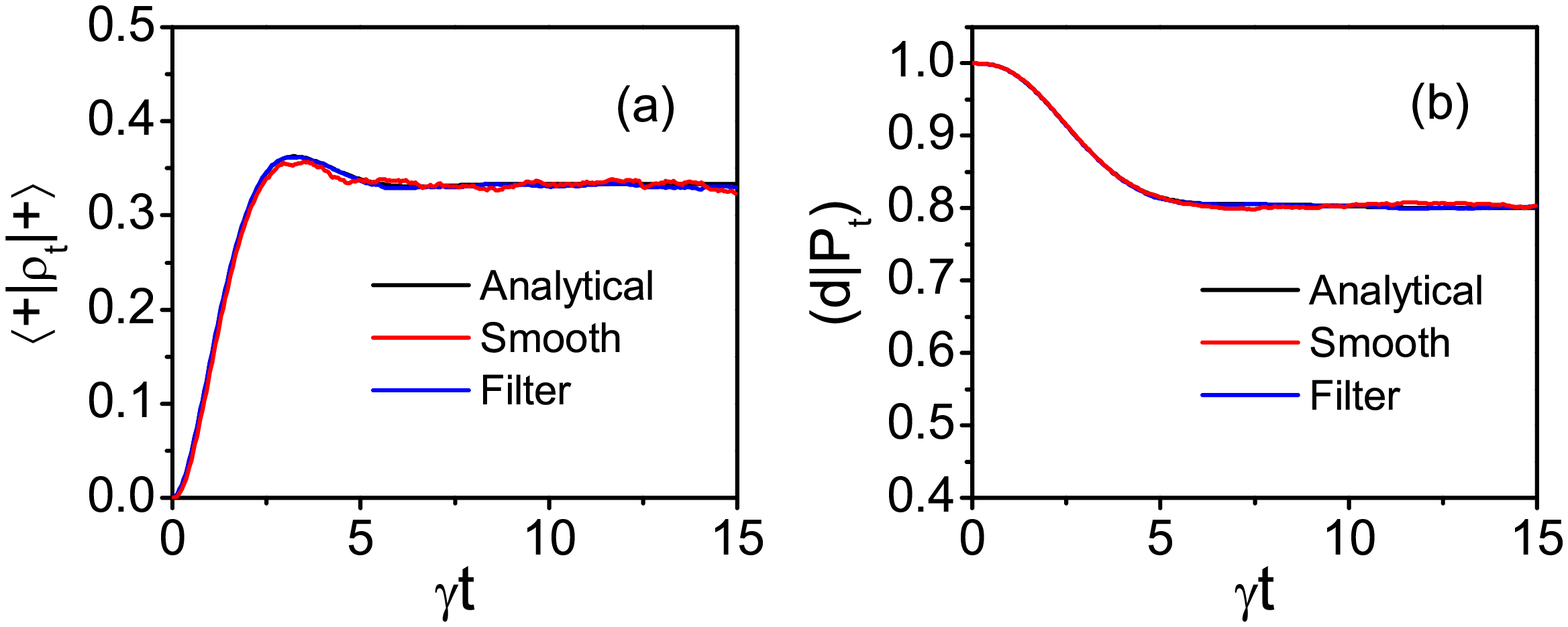}
\caption{Analytical solutions and average over realizations of the filtered
and smoothed states. (a) Upper system population, $\langle +|\protect\rho %
_{t}|+\rangle $ analytical solution, jointly with the smoothed $\protect%
\overleftrightarrow{\langle +|\protect\rho _{t,T}^{\mathrm{st}}|+\rangle }$
and filtered $\protect\overleftarrow{\langle +|\protect\rho _{t}^{\mathrm{st}%
}|+\rangle }$ averaged populations. (b) Classical population, $(d|P_{t})$
analytical solution, jointly with the smoothed $\protect\overleftrightarrow{%
(dP_{t,T}^{\mathrm{st}})}$ and filtered $\protect\overleftarrow{(d|P_{t}^{%
\mathrm{st}})}$ averaged populations. In all cases the averages were
performed with $5\times 10^{5}$ realizations. The parameters are $\Omega /%
\protect\gamma =1$ and $\protect\eta =0.8.$}
\end{figure}
%figura%figura%figura%figura%figurav%figura%figura%figura%figura%figura%figura%figura%figura%figura%figurav%figura%figura%figura%figura%figura
%figura%figura%figura%figura%figurav%figura%figura%figura%figura%figura%figura%figura%figura%figura%figurav%figura%figura%figura%figura%figura

\section{Summary and Conclusions}

An extra class of smoothed quantum state was introduced. It describes a
hybrid quantum-classical arrange conditioned not only on earlier (filtering)
but also later measurements results (smoothing). The joint evolution is
given by an arbitrary hybrid Lindblad rate equation. Hence, mutual influence
between the quantum and classical subsystems is allowed. The measurement
process is performed on the quantum system.

The results relies on a Bayesian analysis, which provides a better
estimation of the\ classical system state [Eq.~(\ref{PastState})], which in
turn lead to an improved estimation\ of the joint state [Eq.~(\ref{Main})].
Partial smoothed states follows by tracing the partner system information
[Eq.~(\ref{PartialSmooth})]. The hybrid smoothed state can be determined
after knowing the initial joint state, the hybrid evolution and the
measurement results. The estimation does not rely on unobserved information
such as for example that provided by measurement processes performed on the
classical subsystem.

By averaging the smoothed state over future measurement results the filtered
state is recovered [Eq.~(\ref{Promedios})]. This property guarantees that a
purer estimation of the joint state is always obtained. Furthermore, and
similarly to the standard quantum jump approach, here the time-irreversible
joint dynamics is recovered after averaging over both past and future
measurement results.

The formalism was applied to a standard fluorescent system monitored by an
inefficient photon detector. This situation was covered by introducing
fictitious classical degrees of freedom (Fig. 1) associated to the imperfect
photon detector. The joint quantum-classical dynamics [Eq.~(\ref{detector})]
leads to the same quantum system dynamics. A significant recovering of the
purity lost due to the inefficient recording process is achieved by taking
into account future measurement results (Fig. 2 and 3). Consistently, the
ensemble averages of both the filtered and smoothed realizations recover the
quantum irreversible system dynamics (Fig. 4).

The present results can be extended an applied in different physical
situations For example, many hybrid measurement channels as well as
application in single-molecule spectroscopy \cite{JPB,OpenSMS} can be
straightforwardly handled by using the developed theoretical formalism.

\section*{Acknowledgments}

This work was supported by CONICET, Argentina.

\appendix

\section{Vectorial quantum jump approach}

Here, we review the quantum jump approach formulated for quantum-classical
hybrid dynamics \cite{JPB} described through a Lindblad rate equation, Eq. (%
\ref{VectorialLindbladRate}).

Given the relation (\ref{LDJ}), the evolution of the joint state can be
rewritten as%
\begin{equation}
\frac{d\left\vert \rho _{t}\right) }{dt}=(\mathcal{\hat{D}}+\mathcal{\hat{J}}%
)\left\vert \rho _{t}\right) .  \label{deterministic}
\end{equation}%
This evolution can be \textquotedblleft unravelled\textquotedblright\ in
terms of measurements trajectories. By solving the previous Markovian
evolution as $|\rho _{t})=e^{\mathcal{\hat{D}}t}\left\vert \rho _{0}\right)
+\int_{0}^{t}e^{\mathcal{\hat{D}}(t-\tau )}\mathcal{\hat{J}}\left\vert \rho
_{\tau }\right) d\tau ,$ after successive iterations, it follows%
\begin{equation}
|\rho _{t})=\mathcal{\hat{G}}(t)|\rho _{0})=\sum_{n=0}^{\infty }\mathcal{%
\hat{G}}_{n}(t)|\rho _{0}).  \label{SerieG}
\end{equation}%
Here, $\mathcal{\hat{G}}_{0}(t)=e^{\mathcal{\hat{D}}t},$ while%
\begin{equation}
\mathcal{\hat{G}}_{n}(t)=\int_{0}^{t}\overleftarrow{dt_{n}}\mathcal{\hat{U}}%
[t,0,\overleftarrow{t_{n}}],
\end{equation}%
where the propagator $\mathcal{\hat{U}}$ is defined by Eq. (\ref{Propa}).
Furthermore, $\overleftarrow{t_{n}}=\{t_{i}\}_{i=1}^{i=n}$ are the
integration variables corresponding to the nested integrals $\int_{0}^{t}%
\overleftarrow{dt_{n}}\equiv \int_{0}^{t}dt_{n}\cdots
\int_{0}^{t_{3}}dt_{2}\int_{0}^{t_{2}}dt_{1}.$

Each contribution in Eq. (\ref{SerieG}) can be rewritten as%
\begin{equation}
\mathcal{\hat{G}}_{n}(t)|\rho _{0})=\int_{0}^{t}\overleftarrow{dt_{n}}P_{t}[%
\overleftarrow{t_{n}}]\frac{\mathcal{\hat{U}}[t,0,\overleftarrow{t_{n}}%
]\left\vert \rho _{0}\right) }{\mathrm{Tr}[(1|\mathcal{\hat{U}}[t,0,%
\overleftarrow{t_{n}}]\left\vert \rho _{0}\right) ]},  \label{Gn}
\end{equation}%
where $P_{t}[\overleftarrow{t_{n}}],$ similarly to Eq. (\ref%
{ProbabilityTrajectory}), is defined as%
\begin{equation}
P_{t}[\overleftarrow{t_{n}}]=\mathrm{Tr}[(1|\mathcal{\hat{U}}[t,0,%
\overleftarrow{t_{n}}]\left\vert \rho _{0}\right) ].  \label{Ptn}
\end{equation}%
This object can be read as the $n$-joint probability density of a
measurement trajectory with $n$-events, each one happening at times $%
\overleftarrow{t_{n}}.$ Hence, from Eq. (\ref{Gn}), the corresponding
conditional stochastic joint state $\left\vert \rho _{t}^{\mathrm{st}%
}\right) $ (associated to the set $\overleftarrow{t_{n}})$ is%
\begin{equation}
\left\vert \rho _{t}^{\mathrm{st}}\right) =\frac{\mathcal{\hat{U}}[t,0,%
\overleftarrow{t_{n}}]\left\vert \rho _{0}\right) }{\mathrm{Tr}[(1|\mathcal{%
\hat{U}}[t,0,\overleftarrow{t_{n}}]\left\vert \rho _{0}\right) ]},
\label{RhoTn}
\end{equation}%
which recovers Eq. (\ref{Filtered}). In this way, Eq. (\ref{SerieG}) can be
read as an \textit{addition over all possible trajectories} with $n$-events
happening at the arbitrary times $\overleftarrow{t_{n}}.$ By construction,
the fulfillment of condition (\ref{AveragingPast}) is guaranteed. Notice
that $P_{t}[\overleftarrow{t_{n}}]$ satisfies the normalization%
\begin{equation}
\sum_{n=0}^{\infty }\int_{0}^{t}\overleftarrow{dt_{n}}P_{t}[\overleftarrow{%
t_{n}}]=1.
\end{equation}

The previous associations are consistent with the definitions of the
measurement transformation Eq. (\ref{EME}) and conditional evolution Eq. (%
\ref{TCond}). In fact, by using the mathematical principle of induction, it
is possible to rewrite Eq. (\ref{RhoTn}) as%
\begin{equation}
\left\vert \rho _{t}^{\mathrm{st}}\right) =\mathcal{\hat{T}}(t,t_{n})%
\mathcal{\hat{M}}\cdots \mathcal{\hat{T}}(t_{2},t_{1})\mathcal{\hat{M}\hat{T}%
}(t_{1},0)|\rho _{0}).  \label{Sucesivo}
\end{equation}%
Therefore, the conditional state can in fact be written as successive
applications of the measurement transformation $\mathcal{\hat{M}},$ while $%
\mathcal{\hat{T}}$ gives the normalized conditional propagation between
measurement events.

Interestingly, by using the mathematical principle of induction it is also
possible to proof that the $n$-joint probability, Eq. (\ref{Ptn}), can be
rewritten as%
\begin{eqnarray}
P_{t}[\overleftarrow{t_{n}}] &=&P_{0}[t,t_{n};\mathcal{\hat{M}}\left\vert
\rho _{t_{n}}^{\mathrm{st}}\right) ]w[t_{n},t_{n-1};\mathcal{\hat{M}}|\rho
_{t_{n-1}}^{\mathrm{st}})]  \notag \\
&&\times \cdots w[t_{2},t_{1};\mathcal{\hat{M}}|\rho _{t_{1}}^{\mathrm{st}%
})]w[t_{1},0;|\rho _{0})].
\end{eqnarray}%
In this expression, for $i\geq 1$%
\begin{equation}
|\rho _{t_{i+1}}^{\mathrm{st}})=\mathcal{\hat{T}}(t_{i+1},t_{i})\mathcal{%
\hat{M}}|\rho _{t_{i}}^{\mathrm{st}}),
\end{equation}%
while $|\rho _{t_{1}}^{\mathrm{st}})=\mathcal{\hat{T}}(t_{1},0)|\rho _{0}).$
The function $w[t,\tau ;|\rho )]$ can be read as a waiting time distribution 
\cite{carmichaelbook}, that is, given that at time $\tau $ the joint state
is $|\rho ),$ it gives the probability density for a interval $t-\tau $
between consecutive measurement events. It reads%
\begin{equation}
w[t,\tau ;|\rho )]\equiv \mathrm{Tr}[(1|\mathcal{\hat{J}}e^{\mathcal{\hat{D}}%
(t-\tau )}|\rho )].  \label{waiting}
\end{equation}%
On the other hand, $P_{0}[t,\tau ;|\rho )]$ is the associated survival
probability%
\begin{equation}
P_{0}[t,\tau ;\left\vert \rho \right) ]=1-\int_{\tau }^{t}w[t^{\prime },\tau
;|\rho )]dt^{\prime },
\end{equation}%
being defined as%
\begin{equation}
P_{0}[t,\tau ;|\rho )]\equiv \mathrm{Tr}[(1|e^{\mathcal{\hat{D}}(t-\tau
)}|\rho )].
\end{equation}

Given an initial condition $|\rho _{0}),$ the dynamics defined by Eq. (\ref%
{Sucesivo}) can be numerically implemented by getting the random measurement
times from the survival probability \cite{JPB}. As in the standard quantum
jump approach it is also possible to write an explicit stochastic
differential equation for the state $|\rho _{t}^{\mathrm{st}}),$ whose
average over realizations recovers the deterministic evolution (\ref%
{deterministic}).

\section{Averaging over future measurements results}

Here, we demonstrate that the average of the joint smoothed state over
future realizations recovers the filtered state, $\overrightarrow{|\rho
_{t,T}^{\mathrm{st}})}=|\rho _{t}^{\mathrm{st}}),$ Eq. (\ref{Promedios}).
Given the expression (\ref{Main}), this is equivalent to demonstrate that
the average of the smoothed conditional probability $P_{T}[R_{t}|%
\overleftrightarrow{t}]$ over measurements performed in the future recovers
the filtered conditional probability $P_{t}[R_{t}|\overleftarrow{t}]=\mathrm{%
Tr}[(R_{t}\left\vert \rho _{t}^{\mathrm{st}}\right) ].$ In an explicit way,
the previous condition can be written as%
\begin{equation}
P_{t}[R_{t}|\overleftarrow{t}]=\int \overrightarrow{dt}P_{T}[R_{t}|%
\overleftarrow{t}\overrightarrow{t}]P_{T}[\overrightarrow{t}|\overleftarrow{t%
}].  \label{triplete}
\end{equation}%
For clarity $P_{T}[R_{t}|\overleftrightarrow{t}]$ was denoted as $%
P_{T}[R_{t}|\overleftarrow{t}\overrightarrow{t}].$ Furthermore, the integral 
$\int \overrightarrow{dt}$ [see Eq. (\ref{Integral}) below] is an addition
over all possible measurement trajectories in $(t,T)$ \textit{given} that we
know one trajectory in $(0,t),$ which is defined by the set of times $%
\overleftarrow{t}.$ In what follows we demonstrate the validity of Eq. (\ref%
{triplete}).

The conditional probability density $P_{T}[\overrightarrow{t}|\overleftarrow{%
t}]$ for the times $\overrightarrow{t}$\ of future measurements given the
past measurement times $\overleftarrow{t}$\ fulfills the Bayes relation%
\begin{equation}
P_{T}[\overleftrightarrow{t}]=P_{T}[\overrightarrow{t}|\overleftarrow{t}%
]P_{t}[\overleftarrow{t}].
\end{equation}%
Here, $P_{t}[\overleftarrow{t}]$ is defined by Eq. (\ref%
{ProbabilityTrajectory}) while $P_{T}[\overleftrightarrow{t}]$ from Eq. (\ref%
{PT}), which lead to%
\begin{equation}
P_{T}[\overrightarrow{t}|\overleftarrow{t}]=\frac{\mathrm{Tr}[(1|\mathcal{%
\hat{U}}[T,t,\overrightarrow{t}]\mathcal{\hat{U}}[t,0,\overleftarrow{t}%
]\left\vert \rho _{0}\right) ]}{\mathrm{Tr}[(1|\mathcal{\hat{U}}[t,0,%
\overleftarrow{t}]\left\vert \rho _{0}\right) ]}.
\end{equation}%
By using this result and Eq. (\ref{PSmoothErre}) for $P_{T}[R_{t}|%
\overleftarrow{t}\overrightarrow{t}],$ Eq.~(\ref{triplete}) becomes%
\begin{equation}
P_{t}[R_{t}|\overleftarrow{t}]=\int \overrightarrow{dt}\frac{\mathrm{Tr}[(1|%
\mathcal{\hat{U}}[T,t,\overrightarrow{t}]|R_{t})(R_{t}|\mathcal{\hat{U}}[t,0,%
\overleftarrow{t}]\left\vert \rho _{0}\right) ]}{\mathrm{Tr}[(1|\mathcal{%
\hat{U}}[t,0,\overleftarrow{t}]\left\vert \rho _{0}\right) ]}.
\label{Parcial}
\end{equation}%
The integral $\int \overrightarrow{dt}$ is an addition over all possible
measurement trajectories in $(t,T).$ Hence, it is given by%
\begin{equation}
\int \overrightarrow{dt}=\sum_{N=0}^{\infty }\int_{t}^{T}\overrightarrow{%
dt_{N}}=\sum_{N=0}^{\infty }\int_{t}^{T}dt_{N}\cdots
\int_{t}^{t_{3}}dt_{2}\int_{t}^{t_{2}}dt_{1},  \label{Integral}
\end{equation}%
where the addition takes into account an arbitrary number of detection
events in the time interval $(t,T).$ By working in a Laplace domain, it is
possible to demonstrate that%
\begin{equation}
\int \overrightarrow{dt}\mathcal{\hat{U}}[T,t,\overrightarrow{t}]=\exp [(T-t)%
\mathcal{\hat{L}}],
\end{equation}%
where $\mathcal{\hat{U}}$ is the propagator (\ref{Propa}) and $\mathcal{\hat{%
L}}$ defines the Lindblad rate equation (\ref{VectorialLindbladRate}). Using
this result and the trace conservation property%
\begin{equation}
\mathrm{Tr}[(1|\exp [t\mathcal{\hat{L}}]|\rho )]=\mathrm{Tr}[(1|\rho )],
\end{equation}%
Eq. (\ref{Parcial}) becomes%
\begin{eqnarray}
P_{t}[R_{t}|\overleftarrow{t}] &=&\frac{\mathrm{Tr}[(R_{t}|\mathcal{\hat{U}}%
[t,0,\overleftarrow{t}]\left\vert \rho _{0}\right) ]}{\mathrm{Tr}[(1|%
\mathcal{\hat{U}}[t,0,\overleftarrow{t}]\left\vert \rho _{0}\right) ]}, \\
&=&\mathrm{Tr}[(R_{t}\left\vert \rho _{t}^{\mathrm{st}}\right) ],
\end{eqnarray}%
where the expression (\ref{Filtered}) was used. The last equality
demonstrates the validity of Eq. (\ref{triplete}) and in consequence also
the validity of Eq. (\ref{Promedios}).

\section{Detection-times}

In an experimental situation, the detection times are determined from the
photon detector. Instead, here they are obtained from the quantum jump
approach. In fact, this formalism not only allows us to defining the
filtered state, but also allows to determining the measurement statistics. A
waiting time distribution \cite{carmichaelbook}\ gives the probability
density for the time interval between consecutive events, Eq. (\ref{waiting}%
).

For a fluorescent system monitored with an inefficient detector $\eta <1$
[Eqs. (\ref{fluor}), (\ref{JDFluor}) and (\ref{MFluor})], by working Eq. (%
\ref{waiting}) in a Laplace domain $[f(u)=\int_{0}^{\infty }dtf(t)e^{-ut}]$
we get $[(t-\tau )\rightarrow u,\ w[t,\tau ;|\rho )]\rightarrow w_{\eta }(u)]
$%
\begin{equation}
w_{\eta }(u)=\frac{\gamma \eta \Omega ^{2}}{u(u+\gamma )(2u+\gamma
)+(2u+\gamma \eta )\Omega ^{2}}.
\end{equation}%
Given that the reseted state (\ref{MFluor}) does not depend on the
(previous) state of the system, $w_{\eta }(u)$ inherits this property
leading to a renewal point process. Exactly the same expression and property
follows from Eq. (\ref{waiting}) calculated over the basis of the Lindblad
rate equation (\ref{detector}) and the vectorial superoperator (\ref%
{Jvectorial}). This feature also demonstrates that the quantum-classical
representation leads to the same quantum system dynamics.

Interestingly, the previous expression can be written as%
\begin{equation}
w_{\eta }(u)=\frac{\eta w_{1}(u)}{1-(1-\eta )w_{1}(u)},
\end{equation}%
where $w_{1}(u)=w_{\eta }(u)|_{\eta =1},$ that is, the waiting time
distribution for perfect detection, $\eta =1.$ By using the geometric series
it follows%
\begin{equation}
w_{\eta }(u)=\eta w_{1}(u)\sum_{n=0}^{\infty }[(1-\eta )w_{1}(u)]^{n}.
\end{equation}%
In this way, $w_{\eta }(u)$ is determined from successive convolution terms,
each one representing a time interval where $n$-fails detection events
happen with probability $(1-\eta )^{n}$ and a detection event happen with
probability $\eta .$ Consequently, one can determine the random events from $%
w_{\eta }(u).$ Equivalently, each event is chosen in agreement with $%
w_{1}(u) $ [perfect detection] and each event is accepted or rejected with
probabilities $\eta $ and $(1-\eta )$ respectively. This last algorithm
recover the expected definition of an inefficient photon detector.


\begin{thebibliography}{99}
\bibitem{breuerbook} H. P. Breuer and F. Petruccione, \textit{The theory of
open quantum systems}, (Oxford University press, 2002).

\bibitem{carmichaelbook} H. J. Carmichael, \textit{An Open Systems Approach
to Quantum Optics}, Lecture Notes in Physics, Vol. M18 (Springer, Berlin,
1993).

\bibitem{plenio} M. B. Plenio and P. L. Knight, The quantum-jump approach to
dissipative dynamics in quantum optics, Rev. Mod. Phys. \textbf{70}, 101
(1998).

\bibitem{milburn} H. M. Wiseman and G. J. Milburn, \textit{Quantum
Measurement and Control} (Cambridge University press, 2010).

\bibitem{recipes} W. H. Press,S. A. Teukolsky,W. T. Vetterling, and B. P.
Flannery, \textit{Numerical Recipes: The Art of Scientific Computing}, 3rd
ed. (Cambridge University Press, New York, 2007).

\bibitem{jaz} A. H. Jazwinski, \textit{Stochatic Processes and Filtering
Theory} (Academic Press, New York, 1970).

\bibitem{tsang} M. Tsang, Time-symmetric quantum theory of smoothing, Phys.
Rev. Lett. \textbf{102}, 250403 (2009).

\bibitem{tsanPRA} M. Tsang, Optimal waveform estimation for classical and
quantum systems via time-symmetric smoothing, Phys. Rev. A \textbf{80},
033840 (2009); M. Tsang, Optimal waveform estimation for classical and
quantum systems via time-symmetric smoothing. II. Applications to atomic
magnetometry and Hardy's paradox, Phys. Rev. A \textbf{81}, 013824 (2010);
M. Tsang, H. M. Wiseman, and C. M. Caves, Fundamental quantum limit to
waveform estimation, Phys. Rev. Lett. \textbf{106}, 090401 (2011).

\bibitem{molmer} S. Gammelmark, B. Julsgaard, and K. M\o lmer, Past Quantum
States of a Monitored System, Phys. Rev. Lett. \textbf{111}, 160401 (2013).

\bibitem{meschede} S. Gammelmark, K. M\o lmer, W. Alt, T. Kampschulte, and
D. Meschede, Hidden Markov model of atomic quantum jump dynamics in an
optically probed cavity, Phys. Rev. A \textbf{89}, 043839 (2014).

\bibitem{murch} D. Tan, S. J. Weber, I. Siddiqi, K. M\o lmer, and K. W.
Murch, Prediction and Retrodiction for a Continuously Monitored
Superconducting Qubit, Phys. Rev. Lett. \textbf{114}, 090403 (2015).

\bibitem{haroche} T. Rybarczyk, B. Peaudecerf, M. Penasa, S. Gerlich, B.
Julsgaard, K. M\o lmer, S. Gleyzes, M. Brune, J. M. Raimond, S. Haroche, and
I. Dotsenko, Forward-backward analysis of the photon-number evolution in a
cavity, Phys. Rev. A \textbf{91}, 062116 (2015).

\bibitem{xu} Q. Xu, E. Greplova, B. Julsgaard, and K. M\o lmer, Correlation
functions and conditioned quantum dynamics in photodetection theory, Phys.
Scr. \textbf{90}, 128004 (2015).

\bibitem{tan} D. Tan, M. Naghiloo, K. M\o lmer, and K. W. Murch, Quantum
smoothing for classical mixtures, Phys. Rev. A \textbf{94}, 050102(R) (2016).

\bibitem{naghi} N. Foroozani, M. Naghiloo, D. Tan, K. M\o lmer and K. W.
Murch, Correlations of the Time Dependent Signal and the State of a
Continuously Monitored Quantum System, Phys. Rev. Lett. \textbf{116}, 110401
(2016).

\bibitem{huard} P. Campagne-Ibarcq, L. Bretheau, E. Flurin, A. Auff\`{e}ves,
F. Mallet, and B. Huard, Observing Interferences between Past and Future
Quantum States in Resonance Fluorescence, Phys. Rev. Lett. \textbf{112},
180402 (2014).

\bibitem{wiseman} I. Guevara and H. Wiseman, Quantum State Smoothing, Phys.
Rev. Lett. \textbf{115}, 180407 (2015).

\bibitem{rate} A. A. Budini, Lindblad rate equations, Phys. Rev. A \textbf{74%
}, 053815 (2006).

\bibitem{dual} A. A. Budini, Operator Correlations and Quantum Regression
Theorem in Non-Markovian Lindblad Rate Equations, J. Stat. Phys. \textbf{131}%
, 51 (2008).

\bibitem{JPB} A. A. Budini, Quantum jumps and photon statistics in
fluorescent systems coupled to classically fluctuating reservoirs, J. Phys.
B \textbf{43}, 115501 (2010).

\bibitem{OpenSMS} A. A. Budini, Open quantum system approach to
single-molecule spectroscopy, Phys. Rev. A \textbf{79}, 043804 (2009).

%\bibitem{waiter} H. J. Carmichael, S. Singh, R. Vyas, and P. R. Rice,
%Photoelectron waiting times and atomic state reduction in resonance
%fluorescence, Phys. Rev. A \textbf{39}, 1200 (1989).

%\bibitem{weak} J. Dressel, M. Malik, F. M. Miatto, A. N. Jordan, and R. W.
%Boyd, Understanding quantum weak values: Basics and applications, Rev. Mod.
%Phys. \textbf{86}, 307 (2014).
\end{thebibliography}
\end{document}